\documentstyle[aps,preprint]{revtex}

\begin{document}

\title{
Separabelized Skyrme Interactions and  Quasiparticle RPA}
\author{A.P. Severyukhin}
\address{Bogoliubov Laboratory of Theoretical Physics, Joint Institute for
Nuclear Research, 141980 Dubna, Moscow region, Russia}
\author{V.V. Voronov}
\address{Bogoliubov Laboratory of Theoretical Physics, Joint Institute for
Nuclear Research, 141980 Dubna, Moscow region, Russia}
\author{Ch. Stoyanov}
\address{Institute for Nuclear Research and Nuclear Energy, boul.
Tzarigradsko Chaussee 72, 1784 Sofia, Bulgaria}
\author{Nguyen Van Giai}
\address{Institut de Physique
Nucl\'eaire, Universit\'e Paris-Sud, F-91406 Orsay Cedex, France}

\maketitle
\begin{abstract}
A finite rank separable approximation for the quasiparticle RPA
with Skyrme interactions is applied to study the low lying
quadrupole and octupole states in some S isotopes and  giant resonances in some
spherical nuclei.
It is shown that characteristics  calculated within the suggested
approach are in a good agreement with available experimental data.
\end{abstract}

\section{Introduction}

The random phase approximation (RPA)\cite{R70,BM75,Schuck,solo} with the self-consistent mean-field
derived with making use of the Gogny's interaction\cite{gogny} or the Skyrme-type interactions\cite{vau72,mig}
is nowadays one of the standard tool to perform nuclear structure calculations.
Many properties of the nuclear collective states can be described successfully
within such models \cite{mig,floc78,doba96,colo1,KG00,colo2,FTTZ00,SFK78}.

Due to the anharmonicity of vibrations there is a
coupling between one-phonon and more complex states \cite{BM75,solo}.
The main difficulty is that the complexity of calculations beyond the standard RPA
increases rapidly with the size of the configuration space and one has to work
within limited spaces. It is well known that using simple separable forces
one can perform calculations of nuclear characteristics in very large configuration
spaces since there is no need to diagonalize matrices whose dimensions grow
with the size of configuration space.
For example, the well-known quasiparticle-phonon model (QPM) \cite{solo}
belongs to such a model. Very detailed predictions can be made by QPM
for nuclei away from closed shells\cite{VS83,VVSS85,gsv}.

That is why a finite rank approximation
for the particle--hole (p-h) interaction resulting from
Skyrme-type forces has been suggested in our previous work \cite {gsv98}.
Thus, the self-consistent mean field can be calculated
in the standard way with the original Skyrme interaction whereas the RPA
solutions would be obtained with the finite rank approximation to the
p-h~matrix elements.
It was found that the finite rank approximation reproduces reasonably well
the dipole and quadrupole strength distributions in Ar isotopes \cite {gsv98}.

Recently, we extended the finite rank approximation for p-h interactions
of Skyrme type to take into account pairing \cite {ssvg02}. We tested
our approach to calculate characteristics of the low-lying
quadrupole and octupole states in some spherical nuclei.
In this paper we apply our approach to study the low lying
quadrupole and octupole states in some S isotopes. Choosing as
examples some spherical nuclei we demonstrate an ability of the method
to describe correctly the strength distributions in a broad excitation energy interval.

\section{Basic formulae and details of calculations }

We start from the effective Skyrme interaction\cite{vau72}
and  use the notation of Ref.\cite{sg81} containing explicit
density dependence and all spin-exchange terms.
The single-particle spectrum is calculated within the HF method.
The continuous part of the single-particle spectrum is
discretized by diagonalizing the HF hamiltonian on
the harmonic oscillator basis\cite{BG77}.
The p-h residual interaction $\tilde V_{res}$
corresponding to the Skyrme force and including both direct and exchange
terms can be obtained as the second derivative of the energy density
functional with respect to the density\cite{ber75}.
Following our previous papers\cite{gsv98,ssvg02} we  simplify $\tilde V_{res}$ by
approximating it by its Landau-Migdal form.
For Skyrme interactions all Landau parameters $F_l, G_l, F^{'}_l, G^{'}_l$
with $l > 1$ are zero. Here, we keep only the $l=0$ terms
in $V_{res}$ and in the coordinate
representation one can write it in the following form:
\begin{eqnarray}
V_{res}({\bf r}_1,{\bf r}_2)=N_0^{-1}\left[ F_0(r_1)+G_0(r_1)
{\bf \sigma}_1{\bf \sigma}_2+(F_0^{^{\prime
}}(r_1)+G_0^{^{\prime }}(r_1){\bf \sigma }_1{\bf \sigma}_2){\bf
\tau }_1{\bf \tau }_2\right] \delta ({\bf r}_1-{\bf r
}_2)  \label{eq2}
\end{eqnarray}

where ${\bf \sigma}_i$ and ${\bf \tau}_i$ are the
spin and isospin operators,
and $N_0 = 2k_Fm^{*}/\pi^2\hbar^2$ with $k_F$ and $m^{*}$ standing for the
Fermi momentum and nucleon effective mass.
The expressions for
$F_0, G_0, F^{'}_0, G^{'}_0$ in terms of the Skyrme force parameters can
be found in Ref.\cite{sg81}. Because of
the density dependence of the interaction the Landau parameters of
Eq.(\ref{eq2}) are functions of the coordinate ${\bf r}$.

The p-h residual interaction can be presented as a sum of
$N$ separable terms.
To illustrate a procedure for making the finite rank approximation
we examine only the contribution of the term  $F_0$.
In what follows we use the second quantized representation
and $V_{res}$ can be written as:

\begin{eqnarray}
\hat V_{res} & = & \frac 12\sum_{1234}V_{1234}:a_1^{+}a_2^{+}a_4 a_3:
\end{eqnarray}

where $a^+_1$ ($a_1$) is the particle creation (annihilation) operator
and $1$ denotes the quantum numbers $(n_1l_1j_1m_1)$,

\begin{eqnarray}
V_{1234} = \int \phi^*_1({\bf r}_1)\phi^*_2({\bf r}_2)
V_{res}({\bf r}_1,{\bf r}_2)\phi_3({\bf r}_1)
\phi_4({\bf r}_2) {\bf dr}_1{\bf dr}_2 ,
\end{eqnarray}
\begin{eqnarray}
V_{1234}=\sum_{JM}\hat J^{-2}
\langle j_1||Y_J||j_3\rangle \langle
j_2||Y_J||j_4\rangle I(j_1j_2j_3j_4)
\times\\\nonumber
(-)^{J+j_3+j_4-M-m_3-m_4}\langle j_1m_1j_3-m_3 \mid J-M\rangle
\langle j_2m_2j_4-m_4\mid JM\rangle.
\end{eqnarray}

In the above equation, $\langle j_1 \vert\vert Y_{J} \vert \vert j_3 \rangle$ is
the reduced matrix element of the spherical harmonics $Y_{J \mu}$,
$\hat J = \sqrt {2J+1}$,
and $I(j_1j_2j_3j_4)$ is the radial integral:
\begin{eqnarray}
I(j_1j_2j_3j_4)=N_0^{-1}\int_0^\infty  F_0(r)
u_{j_1}(r)u_{j_2}(r)u_{j_3}(r)u_{j_4}(r)\frac{dr}{r^2},
\end{eqnarray}

where ${\LARGE u(r)}$ is the radial part of the HF single-particle wavefunction.
As it is shown in \cite{gsv98,ssvg02} the radial integrals can be calculated
accurately by choosing a large enough cutoff radius $R$
and using a $N$-point integration Gauss formula with abscissas ${r_k}$ and weights ${w_k}$.
\begin{eqnarray}
I(j_1j_2j_3j_4)\simeq N_0^{-1}\frac{R}{2}\sum_{k=1}^{N}
{\frac{w_kF_0(r_k)}{r_k^2}}
u_{j_1}(r_k)u_{j_2}(r_k)u_{j_3}(r_k)u_{j_4}(r_k)
\end{eqnarray}

So we employ the hamiltonian
including an average nuclear HF field,
pairing interactions, the isoscalar and
isovector particle--hole residual forces
in the finite rank separable form \cite{ssvg02}.
This hamiltonian has the same form as the QPM hamiltonian
with $N$ separable terms \cite{solo,sol89},
but in contrast to the QPM all parameters
of this hamiltonian are expressed through parameters of the
Skyrme forces.

In what follows we work in the quasiparticle  representation defined by
the canonical Bogoliubov transformation:
\begin{equation}
a_{jm}^{+}\,=\,u_j\alpha _{jm}^{+}\,+\,(-1)^{j-m}v_j\alpha _{j-m}.
\label{B}
\end{equation}
The single-particle states are specified by the
quantum numbers $(jm)$
The quasiparticle energies,
the chemical potentials, the energy gap and the coefficients
$u$,$v$ of the  Bogoliubov transformations
(\ref{B}) are determined
from  the BCS equations.

We introduce the phonon creation operators
\begin{equation}
Q_{\lambda \mu i}^{+}\,=\,\frac 12\sum_{jj^{^{\prime }}}\left( X
_{jj^{^{\prime }}}^{\lambda i}\,A^{+}(jj^{^{\prime }};\lambda \mu
)-(-1)^{\lambda -\mu }Y _{jj^{^{\prime }}}^{\lambda i}\,A(jj^{^{\prime
}};\lambda -\mu )\right)
\end{equation}

where
\begin{equation}
A^{+}(jj^{^{\prime }};\lambda \mu )\,=\,\sum_{mm^{^{\prime }}}\langle
jmj^{^{\prime }}m^{^{\prime }}\mid \lambda \mu \rangle \alpha
_{jm}^{+}\alpha _{j^{^{\prime }}m^{^{\prime }}}^{+}.
\end{equation}

The index $\lambda $ denotes total angular momentum and $\mu $ is
its z-projection in the laboratory system.
One assumes that the QRPA ground state  is the phonon vacuum
$\mid 0\rangle $,\\ i.e. $Q_{\lambda \mu i}\mid 0\rangle\,=0$.
We define the excited states for this approximation by
$Q_{\lambda\mu i}^{+}\mid0\rangle$.

Making use of the linearized equation-of-motion approach \cite{R70}
one can derive the QRPA equations \cite{Schuck,solo}:

\begin{equation}
\label{eq14}
\left(
\begin{tabular}{ll}
${\cal A}$ & ${\cal B}$ \\
${- \cal B}$ & ${- \cal A}$%
\end{tabular}
\right) \left(
\begin{tabular}{l}
$ X $ \\
$ Y $%
\end{tabular}
\right) =w \left(
\begin{tabular}{l}
$ X $ \\
$ Y $%
\end{tabular}
\right).
\end{equation}

In QRPA problems there appear two types of interaction matrix elements,
the matrix related to forward-going graphs
${\cal A}^{(\lambda)}_{(j_1j_1^{\prime})(j_2j_2^{\prime})}$ and the
matrix related to backward-going graphs
${\cal B}^{(\lambda)}_{(j_1j_1^{\prime})(j_2j_2^{\prime})}$.
Solutions of this set of linear equations yield the eigen-energies
and the amplitudes $X,Y$ of the excited states.
A dimension of the matrixes ${\cal A}, {\cal B}$ is a space size of
the two-quasiparticle configurations. Expressions for ${\cal A}, {\cal B}$
and $X,Y$ are given in \cite{ssvg02}.

Using the finite rank approximation we need to invert a matrix having
a dimension $4N \times 4N$ independently of the configuration space size.
One can find a prescription how to solve the system (\ref{eq14}) within
our approach in  \cite{gsv98,ssvg02}.
The QRPA equations in the QPM \cite{solo,sol89} have the same form as
the equations derived within our approach\cite{gsv98,ssvg02},
but the single-particle spectrum and
parameters of the p-h residual interaction are
calculated making use of the Skyrme forces.

In this work we use
the standard parametrization SIII \cite{be75} of the Skyrme force.
Spherical symmetry is assumed for
the HF ground states.
It is well
known \cite{KG00,colo2} that the constant gap approximation leads to
an overestimating of occupation probabilities for subshells that are far
from the Fermi level and it is necessary to introduce a cut-off in the
single-particle space. Above this cut-off subshells don't participate in
the pairing effect. In our calculations we choose the BCS subspace
to include all subshells lying below 5 MeV.
The pairing constants are fixed to reproduce the odd-even mass
difference of neighboring nuclei.
In order to perform RPA calculations, the single-particle continuum is
discretized \cite{BG77} by diagonalizing the HF hamiltonian on a basis
of twelve
harmonic oscillator shells and cutting off the single-particle spectra at the
energy of 160 MeV. This is sufficient to exhaust practically all the
energy-weighted sum rule.
Our investigations \cite{ssvg02} enable us to conclude that $N$=45 is enough for
multipolarities $\lambda \le 3 $ in nuclei with $A\le 208$.
Increasing $N$, for example, up to $N$=60 in $^{208}$Pb changes results
for energies
and transition probabilities not more than by 1\%, so all calculations
in what follows have been done with  $N$=45.  Our calculations show
that, for the normal parity states one can neglect the spin-multipole
interactions  as a rule and this reduces by a factor 2
the total matrix dimension. For example, for the octupole
excitations in $^{206}$Pb \cite{ssvg02} we need to invert a matrix having
a dimension 2N=90 instead of diagonalizing a $1376\times1376$ matrix as
it would be without the finite rank approximation.
For light nuclei the reduction of matrix dimensions
due to the finite rank approximation is 3 or 4.
So, for heavy nuclei our approach gives a large gain in comparison
with an exact diagonalization.
It is worth to point out that after
solving the RPA problem with a separable interaction, to take into
account
the coupling with two-phonon configurations demands to diagonalize a
matrix having a size that does not exceed 40 for the giant resonance
calculations in heavy nuclei whereas one would need to diagonalize a matrix
with a dimension of the order of a few thousands at least for a
non-separable case.

\section{Results of calculations}

As a first example we examine  the $2^+_1$ and $3^-_1$ state energies and
transition probabilities in some S isotopes.
The results of our calculations for the energies and B(E2)-values and
the experimental data \cite{Ram01} are shown in Table 1.
One can see that there is a rather good agreement with
experimental data. Results of our calculations for S isotopes
are close to those of QRPA with Skyrme forces \cite{Khan01}.
The evolution of the B(E2)-values in the S isotopes
demonstrates clearly the pairing effects. The experimental and
calculated B(E2)-values in $^{36}$S are two times less than those in
$^{34,38}$S. The neutron shell closure leads to the vanishing of
the neutron pairing  and a reduction of the proton gap. As a result
there is a remarkable reduction of the E2 transition probability in
$^{36}$S. Some overestimate of the energies in $^{34,38}$S indicates
that there is room for two-phonon effects.
The study of the influence of two-phonon configurations
on properties of the low-lying states within our approach
is in progress now.

Results of our calculations for the $3^-_1$ energies and the transition
probabilities B(E3) are compared with experimental data \cite{Sp02} in
Table 2. Generally there is a good agreement between theory
and experiment.

An additional information  about the structure of the first $2^+, 3^-$
states can be extracted by looking at the ratio of the multipole
transition matrix elements $M_n/M_p$ that depend on the relative
contributions of the proton and neutron configurations. In the framework
of the collective model for isoscalar excitations this ratio is
equal to $M_n/M_p=N/Z$ and any deviation from this value can indicate
an isovector character of the state. The $M_n/M_p$ ratio can be determined
experimentally by using different external probes \cite{Ber83,Ken92,Jew99}.
Recently \cite{Khan01}, QRPA calculations of
the $M_n/M_p$ ratios for the $2^+_1$ states in some S
isotopes have been done. The predicted results are in good
agreement with experimental data \cite {Khan01}. Our calculated values of
the $M_n/M_p$ ratios for the $2^+_1$ and $3^-_1$ states are shown in Tables
1 and 2, respectively.
Our results support the conclusions of Refs. \cite{Khan01} about the
isovector character of the $2^+_1$ states in $^{36}$S.
As one can see from Table 2  our calculations predict that
the $M_n/M_p$ ratios for the $3^-_1$ states are rather close to $N/Z$,
thus indicating their isoscalar character.

To test our approach for high lying states we examine the dipole strength
distributions.
The calculated dipole strength distributions (GDR)
in $^{36}$Ar, $^{112}$Sn and $^{208}$Pb are displayed in Fig. 1.
For the energy centroids $(m_1/m_0)$ we get 19.9 MeV, 15.8 MeV and
12.7 MeV in $^{36}$Ar, $^{112}$Sn and $^{208}$Pb respectively.
The calculated energy centroid for $^{208}$Pb is
in a satisfactory agreement with the experimental value \cite{cenpb208} (13.4 MeV).
The values of energy centroids for $^{36}$Ar and $^{112}$Sn are rather
close to the empirical systematics \cite{VW87} $E_c=31.2 A^{-1/3} + 20.6 A^{-1/6}$ MeV.
For $^{36}$Ar  the QRPA gives results that are very similar to our previous calculations
with the particle-hole RPA \cite{gsv98} because the influence of pairing on the giant
resonance properties is weak. It is worth to mention that experimental data for the giant
resonances in light nuclei are very scarce.

The octupole strength distribution in $^{208}$Pb is rather well studied
in many experiments \cite{Fuj,HEOR2}. The calculated octupole strength
distribution up to the excitation energy 35 MeV is shown in Fig. 2.
According to experimental data \cite{Fuj} for the $3^-_1$ state in $^{208}$Pb
the excitation energy equals to $E_x=2.62$ MeV and the energy-weighted sum rule
(EWSR) is exhausted by 20.4\% that can be compared with the calculated values
$E_x=2.66$ MeV and EWSR=21\%.
For the low-energy octupole resonance below 7.5 MeV our calculation gives
the centroid energy $E_c=5.96$MeV and EWSR=12\% and experimental values
are 5.4 MeV and 15.2\% accordingly.
For the high-energy octupole resonance we get values $E_c=20.9$ MeV
and EWSR=61\% that are in a good agreement with
experimental findings $E_c=20.5\pm1$ MeV
and EWSR=$75\pm15$\% \cite{HEOR2}.
One can conclude that present calculations reproduce correctly not only
the $3^-_1$
characteristics, but the whole octupole strength distribution in $^{208}$Pb.

\section{Conclusion}

A finite rank separable approximation for the QRPA
calculations with Skyrme interactions that was proposed in
our previous work is applied to study the evolution of dipole,
quadrupole and octupole excitations  in several nuclei.
It is shown that the suggested approach enables
one to reduce remarkably the dimensions of the matrices that
must be inverted to perform structure calculations in very large
configuration spaces.

As an illustration of the method we have calculated the
energies and transition probabilities of the $1^-$, $2^+$ and $3^-$
states in some S, Ar, Sn and Pb isotopes.  The  calculated
values are very close to those that were
calculated in QRPA with the full Skyrme interactions.
They are in an agreement with available experimental data.

\section{Acknowledgments}

A.P.S. and V.V.V. thank the hospitality of IPN-Orsay where a part
of this work was done. This work is partly supported
by INTAS Fellowship grant for Young Scientists
(Fellowship Reference N YSF 2002-70),
by IN2P3-JINR agreement
and by the Bulgarian Science Foundation (contract Ph-801).


\begin{table}[]
\caption[]{Energies, B(E2)-values and  $(M_n/M_p)/(N/Z)$ ratios for up-transitions
to the first $2^{+}$ states}
\begin{center}
\begin{tabular}{ccccccc}
Nucleus    & \multicolumn{2}{c} {Energy} & \multicolumn{2}{c} {B(E2$\uparrow$)} & \multicolumn{2}{c} {$(M_n/M_p)/(N/Z)$} \\
           & \multicolumn{2}{c} {(MeV)}  & \multicolumn{2}{c} {(e$^2$fm$^4$)}   & \multicolumn{2}{c} {}                  \\
           & Exp. & Theory               & Exp.         & Theory                & Exp.          & Theory                  \\
\hline
  $^{32}$S & 2.23 & 3.34                 & 300$\pm$13   &    340                &  0.94$\pm$0.16& 0.92                    \\
  $^{34}$S & 2.13 & 2.48                 & 212$\pm$12   &    290                &  0.85$\pm$0.23& 0.87                    \\
  $^{36}$S & 3.29 & 2.33                 & 104$\pm$28   &    130                &  0.65$\pm$0.18& 0.40                    \\
  $^{38}$S & 1.29 & 1.55                 & 235$\pm$30   &    300                &  1.09$\pm$0.29& 0.73                    \\
\end{tabular}
\end{center}
\end{table}

\begin{table} []
\caption[]{Energies, B(E3)-values and  $(M_n/M_p)/(N/Z)$ ratios for up-transitions
to the first $3^{-}$ states}
\begin{center}
\begin{tabular}{ccccccc}
Nucleus    & \multicolumn{2}{c} {Energy}  & \multicolumn{2}{c} {B(E3$\uparrow$)}&  {$(M_n/M_p)/(N/Z)$}    \\
           & \multicolumn{2}{c} {(MeV)}   & \multicolumn{2}{c} {(e$^2$fm$^6$)}  &                         \\
           & Exp. & Theory                & Exp.               & Theory         & Theory                  \\
\hline
$^{32}$S   & 5.01 & 7.37                  &12700$\pm$2000      &    8900         & 0.89                    \\
$^{34}$S   & 4.62 & 5.66                  & 8000$\pm$2000      &    8500         & 1.06                    \\
$^{36}$S   & 4.19 & 3.86                  & 8000$\pm$3000      &    7200         & 1.15                    \\
$^{38}$S   & --   & 5.68                  & --                 &    6200         & 1.01                    \\
\end{tabular}
\end{center}
\end{table}

\begin{figure}[]
\caption {Strength distributions of the GDR in $^{36}$Ar, $^{112}$Sn and $^{208}$Pb}
\end{figure}

\begin{figure}[]
\caption {The octupole strength distribution in $^{208}$Pb}
\end{figure}

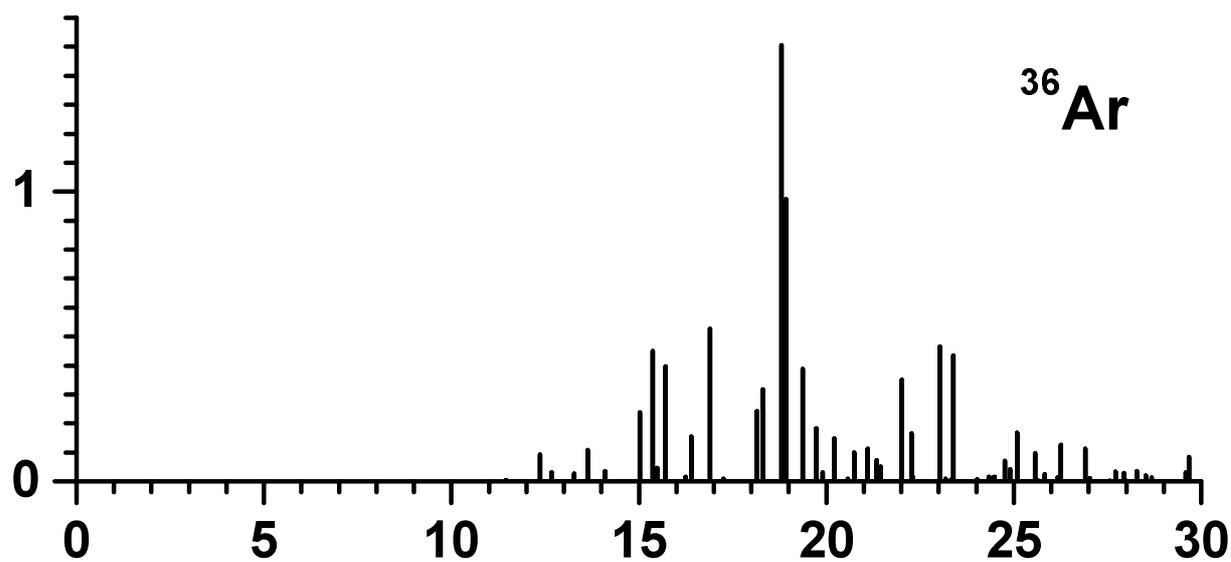
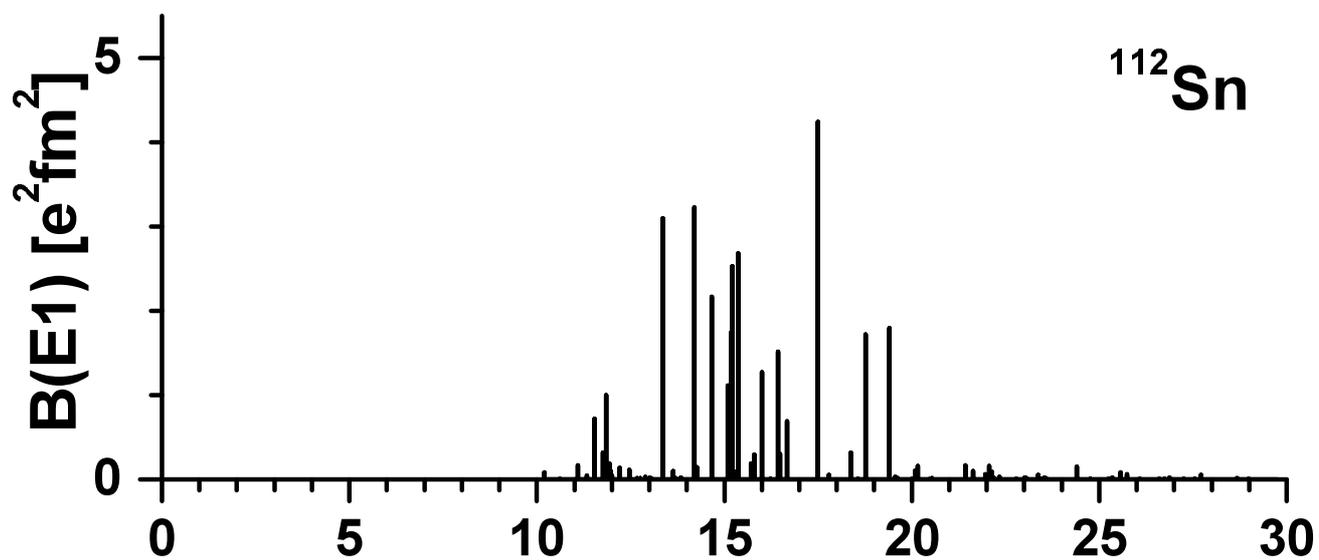
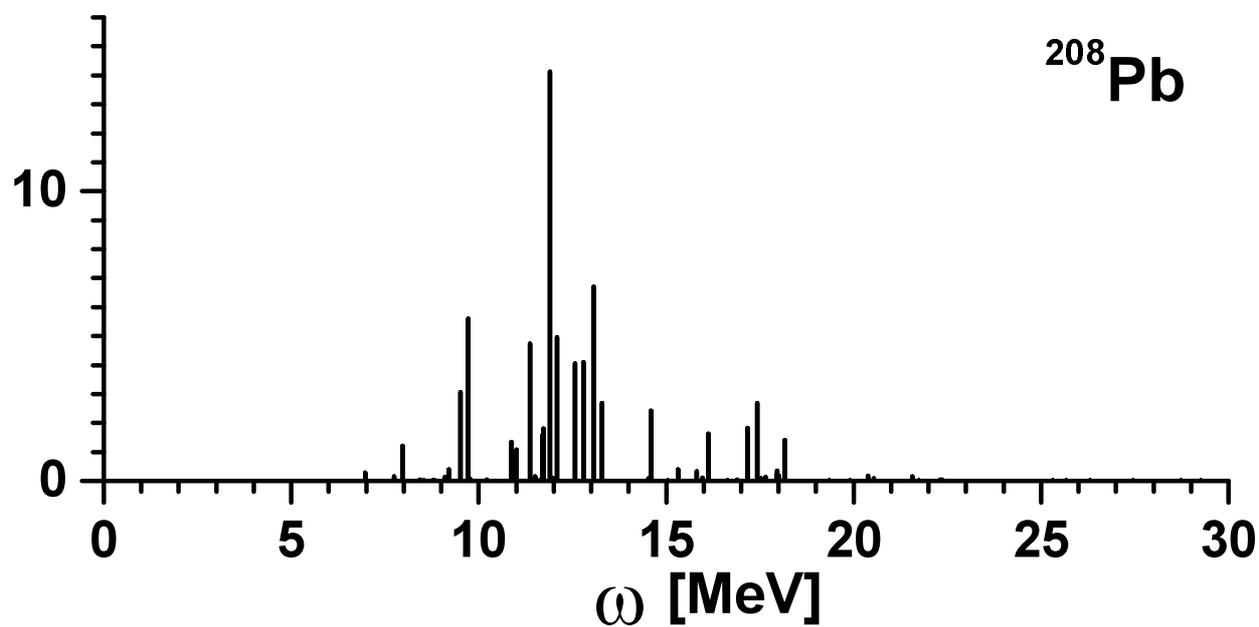
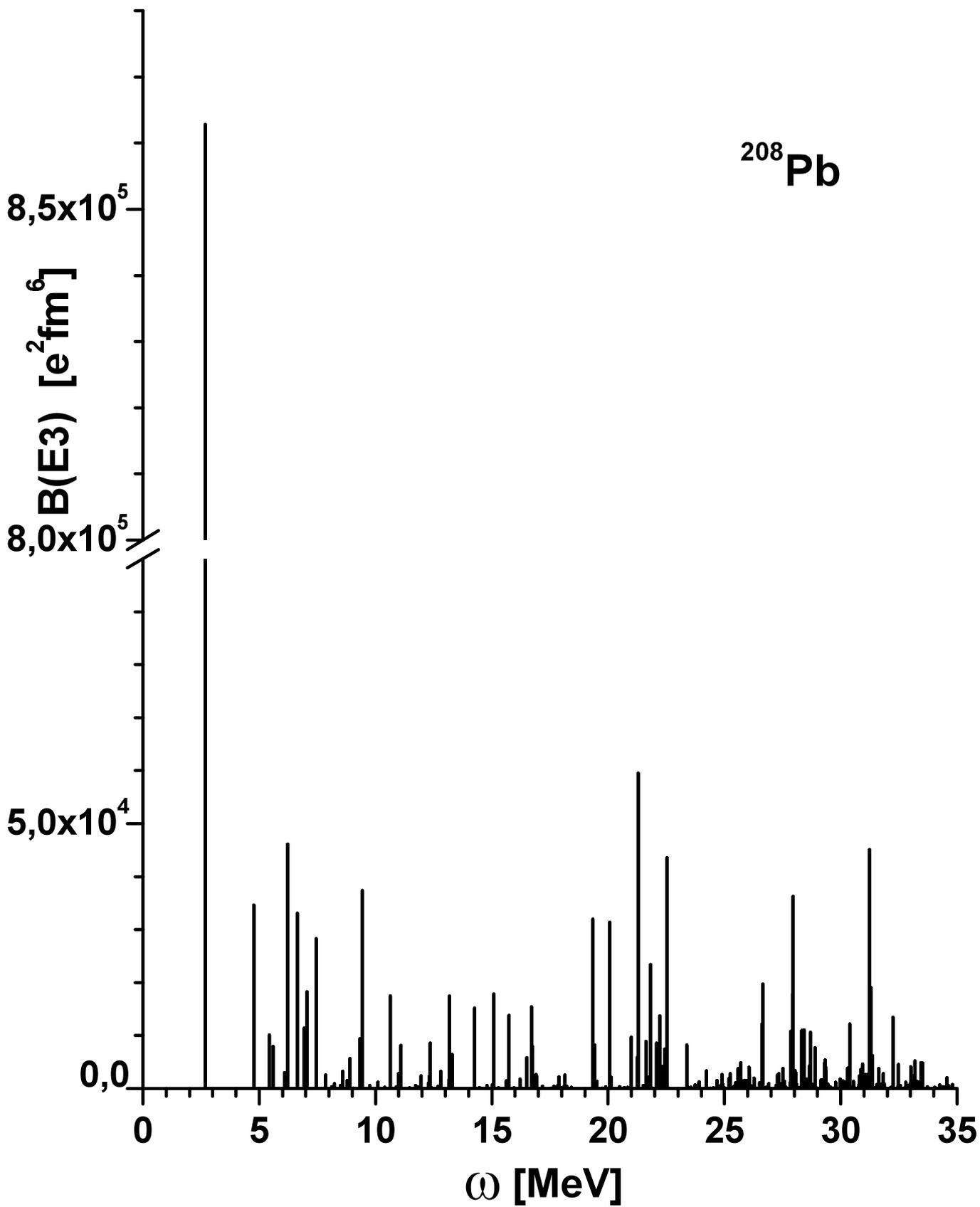
\end{document}